\documentclass[twocolumn]{aastex631}
\usepackage{natbib}

\newcommand{\LIGOlabMIT}{\affiliation{LIGO Laboratory, Massachusetts Institute of Technology, Cambridge, MA 02139, USA}}
\newcommand{\MKI}{\affiliation{Department of Physics and Kavli Institute for Astrophysics and Space Research, \\Massachusetts Institute of Technology, Cambridge, MA 02139, USA}}
\newcommand{\CITA}{\affiliation{Canadian Institute for Theoretical Astrophysics, University of Toronto, Toronto, ON M5S 3H8, Canada}}
\newcommand{\LSU}{\affiliation{Department of Physics \& Astronomy, Louisiana State University, Baton Rouge, LA 70803, USA}}

\begin{document}
\defcitealias{Biscoveanu:2022iue}{SB22}
\title{An observational upper limit on the rate of gamma-ray bursts\\ with neutron star-black hole merger progenitors}

\correspondingauthor{Sylvia Biscoveanu}
\email{sbisco@mit.edu}
\author[0000-0001-7616-7366]{Sylvia Biscoveanu} \MKI \LIGOlabMIT
\author[0000-0002-2942-3379]{Eric Burns} \LSU
\author[0000-0002-8457-1964]{Philippe Landry} \CITA
\author[0000-0003-2700-0767]{Salvatore Vitale} \MKI \LIGOlabMIT

\begin{abstract}

Compact-object binary mergers consisting of one neutron star and one black hole (NSBHs) have long been considered promising progenitors for gamma-ray bursts, whose central engine remains poorly understood. Using gravitational-wave constraints on the population-level NSBH mass and spin distributions we find that at most $20~\mathrm{Gpc}^{-3}\mathrm{yr}^{-1}$ of gamma-ray bursts in the local universe can have NSBH progenitors.

\end{abstract}

\section{Introduction}
\label{sec:intro}
Gamma-ray bursts (GRBs) are among the most energetic electromagnetic explosions in the universe, but the physical mechanism powering the burst central engine remains highly uncertain.
The association of GRB170817A with the binary neutron star merger GW170817 confirmed that at least some short GRBs have compact-object binary progenitors~\citep{LIGOScientific:2017zic}. Neutron star-black hole (NSBH) mergers are another class of binary that may be accompanied by a GRB if the neutron star is tidally disrupted outside the black hole's innermost stable circular orbit, leaving sufficient remnant mass to form an accretion disk that can power the GRB jet~\citep[e.g.,][]{Narayan:1992iy}.

The third observing run (O3) of the LIGO-Virgo-Kagra (LVK) gravitational-wave interferometers~\citep{Aasi2015,AcerneseAgathos2015,AkutsuAndo2021} included the detection of four NSBH candidate events with false alarm rate $\mathrm{FAR} < 1~\mathrm{yr}^{-1}$~\citep{LIGOScientific:2021djp}. Using these four events, we placed constraints on the population-level distributions of the masses and spins of the compact objects in NSBH mergers in \cite{Biscoveanu:2022iue}, henceforth \citetalias{Biscoveanu:2022iue}. The probability of neutron star tidal disruption and hence the remnant mass left after the merger depend on the mass ratio between the neutron star and black hole, the neutron star equation of state, and on the black hole spin aligned to the orbital angular momentum~\citep{Foucart:2018rjc}. We used the population-level mass and spin distributions to place an observational constraint on the fraction of NSBHs detectable in gravitational waves that may be electromagnetically bright by enforcing a minimum remnant mass required for a counterpart. Here, we extend this analysis to place a constraint on the local merger rate of NSBHs that may be electromagnetically bright. This serves as a proxy for the total (beaming-corrected) rate of GRBs with NSBH progenitors.

\section{Methods}
\begin{figure*}
\centering
\includegraphics[width=\columnwidth]{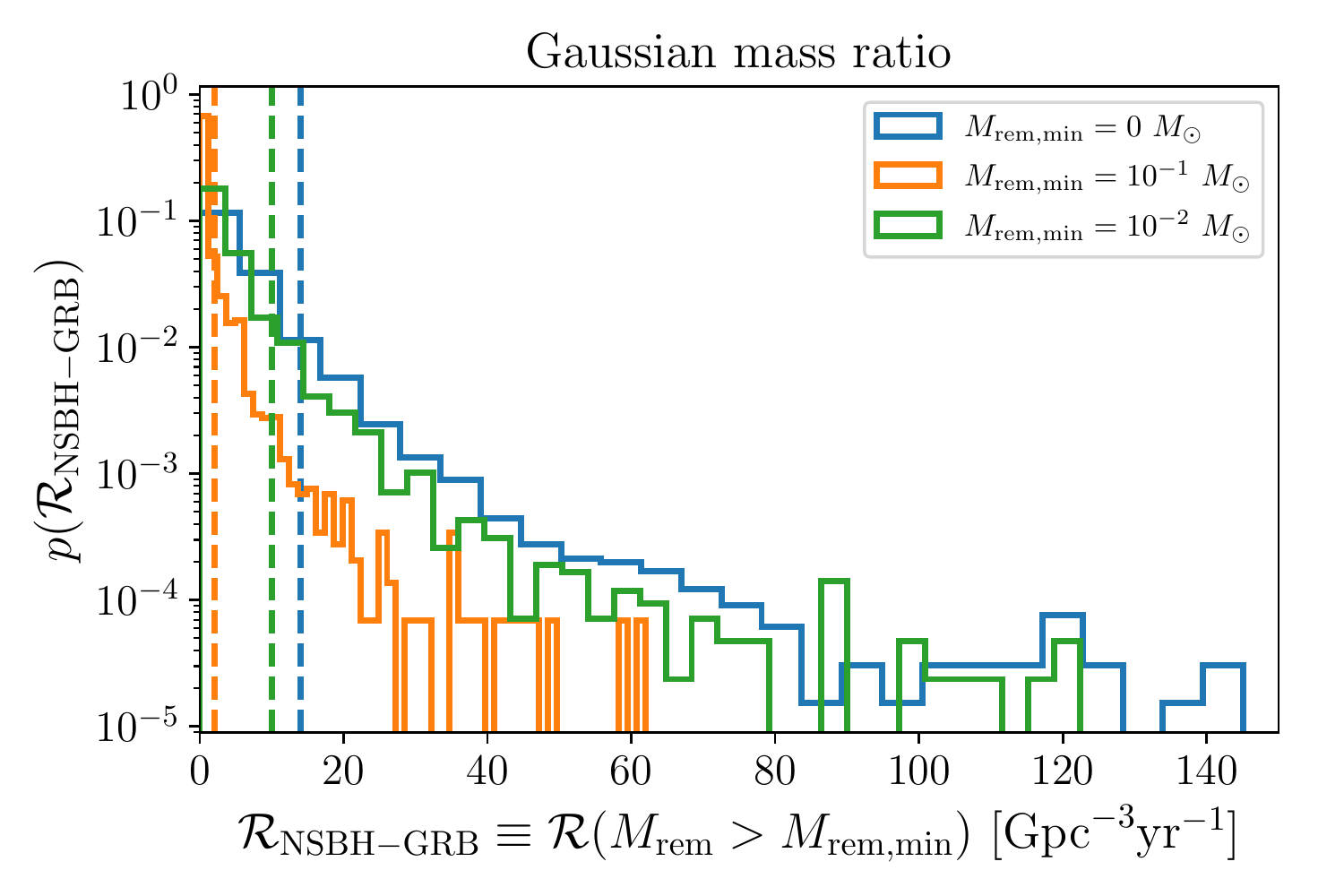}
\includegraphics[width=\columnwidth]{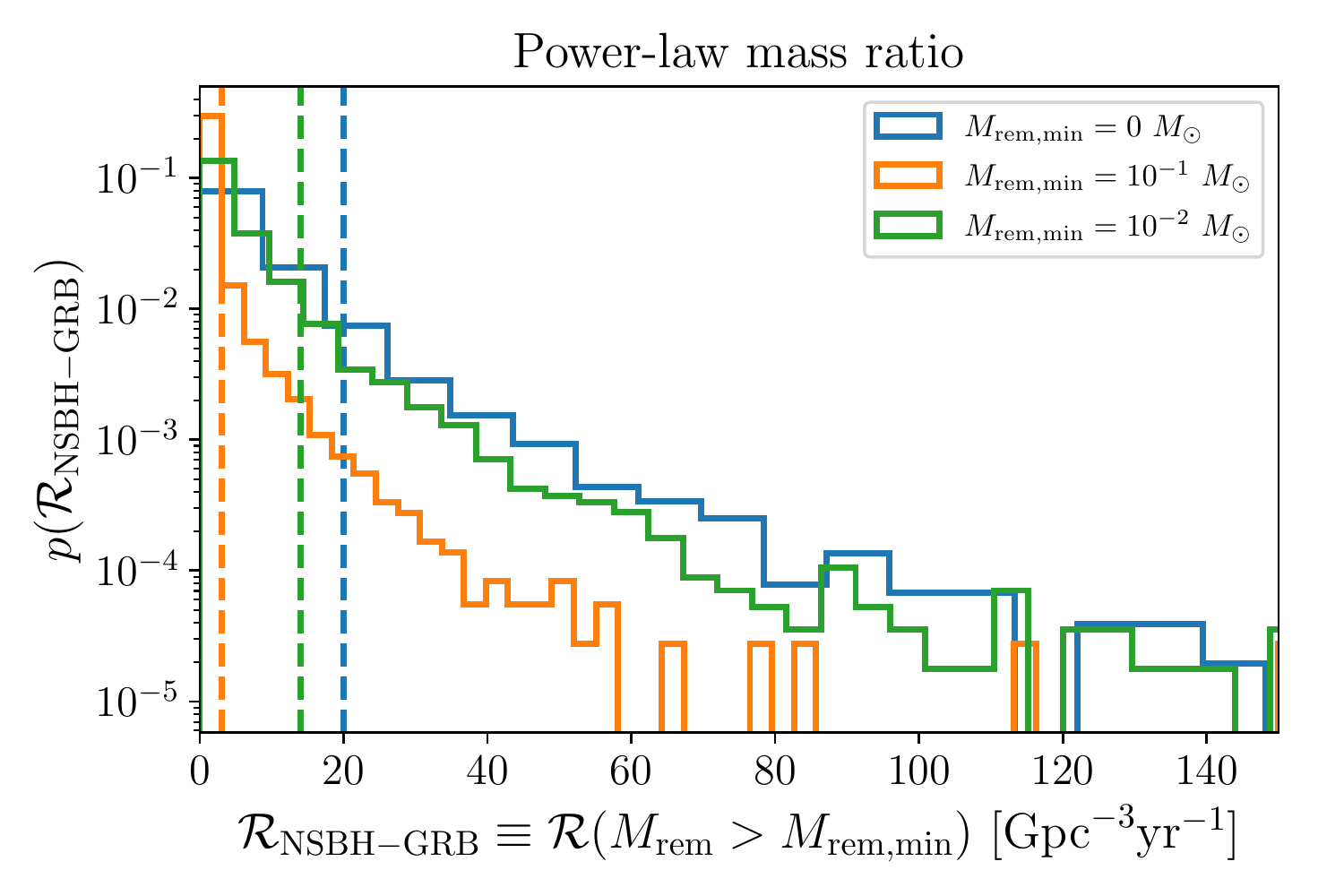}
\caption{Posterior probability distributions on the local rate of GRBs with NSBH progenitors under the Gaussian (left) and power-law (right) mass ratio models. The colors indicate three different values of the minimum remnant mass that we require for a NSBH merger to potentially power a GRB. The dashed vertical lines indicate the 90\textsuperscript{th} percentile for each value of the minimum remnant mass.
\label{fig:embright_total}}
\end{figure*}

We use the binary mass and spin distributions for the NSBH population inferred in \citetalias{Biscoveanu:2022iue}~\citep{biscoveanu_sylvia_2022_6795223} based on hierarchical analysis of the four NSBH candidates from O3. We fit the black hole mass distribution with a truncated power law, the black hole spin magnitude with a Beta distribution, and the mass ratio distribution with either a Gaussian or another power-law. %
We assume that the black hole tilt distribution is isotropic. See \citetalias{Biscoveanu:2022iue} for more details of the hierarchical analysis including priors, distribution functional forms, sampler settings, and selection effect treatment.

We then draw samples in the binary parameters from the inferred population-level distributions. The number of samples drawn is determined by the merger rate at $z=0$ inferred simultaneously with the population-level distributions. We marginalize over the uncertainty in the population-level distributions and in the neutron star equation of state (EoS) using the constraints obtained in \cite{Legred:2021hdx} based on observations of binary neutron star mergers and pulsars. For each population-level distribution considered, we enforce that the maximum neutron star mass should be below the maximum mass of a rigidly rotating neutron star supported by the associated EoS. The neutron star spin magnitude is then drawn uniformly up to the breakup spin supported by that EoS. 

Given a complete set of binary parameters %
we then calculate the remnant mass that would be left outside the black hole innermost stable circular orbit following the merger using the fitting formula from \cite{Foucart:2018rjc}. If the remnant mass is above a threshold, we consider that sample to be a possible GRB progenitor. Because of the considerable uncertainty in the remnant mass required to power a counterpart like a GRB, we consider three different values of the remnant mass threshold, $M_{\mathrm{rem,min}}=0, 0.01, 0.1~M_{\odot}$. %
We then count the total number of potential GRB progenitors among our population of merging NSBHs, defining the maximum local rate of GRBs with NSBH progenitors as $\mathcal{R}_{\mathrm{NSBH-GRB}} \equiv N(M_{\mathrm{rem}} > M_{\mathrm{rem, min}})$.

\section{Results and Discussion}
In Fig.~\ref{fig:embright_total}, we show the posterior probability distributions on $\mathcal{R}_{\mathrm{NSBH-GRB}}$ under both the Gaussian and power-law distributions for the mass ratio. We find an upper limit of $20~\mathrm{Gpc}^{-3}\mathrm{yr}^{-1}$ ($14~\mathrm{Gpc}^{-3}\mathrm{yr}^{-1}$) under the Gaussian (power-law) mass ratio model at 90\% credibility. This is consistent with previous constraints conditioned on the observed rate of short GRBs~\citep{Sarin:2022cmu}. While the reported astrophysical beaming-corrected rate of short GRBs is highly uncertain, ranging from $\mathcal{O}(10)-\mathcal{O}(1000)~\mathrm{Gpc}^{-3}\mathrm{yr}^{-1}$~(see \citealt{Mandel:2021smh} for a review), the upper limit we find for GRBs with NSBH progenitors represents a small fraction of all short GRBs even for the lowest rates. 

The recent discovery of a kilonova accompanying a long GRB suggests that some fraction of these transients may also have a compact-object merger origin~\citep{Troja:2022yya, Rastinejad:2022zbg}. In this case NSBHs would represent a subdominant progenitor class for both long and short GRBs. While this conclusion depends to some extent on how the population of NSBH candidates is chosen, 
in \citetalias{Biscoveanu:2022iue} we showed that the inferred mass and spin distributions---and hence the electromagnetically-bright fraction---do not change substantially when the observed population is restricted to only the two highest-significance events.%

The electromagnetically bright fraction may also change if we had assumed that the black hole spin orientations are aligned to the orbital angular momentum---as suggested if NSBHs form via isolated binary evolution~\citep[e.g.,][]{Broekgaarden:2021iew}---rather than isotropically distributed. The effect is likely to be small since the spin distribution inferred in \citetalias{Biscoveanu:2022iue} prefers small magnitudes. However, only individual-event parameter posteriors  obtained with an isotropic spin distribution are publicly available for all four events we consider, so a comparison to the aligned-spin case is outside the scope of this work. 
The results we find here support the idea that neutron star-black hole mergers are uncommon multimessenger sources. 

%\begin{acknowledgments}
\textit{Acknowledgements. }
S.B. acknowledges support of the National Science Foundation and the LIGO Laboratory. LIGO was constructed by the California Institute of Technology and Massachusetts Institute of Technology with funding from the National Science Foundation and operates under cooperative agreement PHY-0757058. S.B. is also supported by the NSF Graduate Research Fellowship under Grant No. DGE-1122374. 
S.V. is also supported by NSF PHY-2045740. P.L. is supported by the Natural Sciences \& Engineering Research Council of Canada (NSERC). 
The authors are grateful for computational resources provided by the LIGO Lab and supported by NSF Grants PHY-0757058 and PHY-0823459. This note carries LIGO document number LIGO-P2300187.
%\end{acknowledgments}

\bibliography{nsbh_note}{}
\bibliographystyle{aasjournal}

\end{document}